# Light trapping in solar cells at the extreme coupling limit

Ali Naqavi,[1,2,*] Franz-Josef Haug,[1] Corsin Battaglia,[1] Hans Peter Herzig,[2] and Christophe Ballif [1]

[1]*Photovoltaics and Thin Film Electronics Laboratory, Institute of Microengineering (IMT), Ecole Polytechnique Fédérale de Lausanne (EPFL), Rue A.-L. Breguet 2, 2000 Neuchâtel, Switzerland*

[2]*Optics & Photonics Technology Laboratory, Ecole Polytechnique Fédérale de Lausanne (EPFL), Rue A.-L. Breguet 2, 2000 Neuchâtel, Switzerland*

[*]*Corresponding author: ali.naqavi@epfl.ch*

We calculate the maximal absorption enhancement obtainable by guided mode excitation in a weakly absorbing dielectric slab over wide wavelength ranges. The slab mimics thin film silicon solar cells in the low absorption regime. We consider simultaneously wavelength-scale periodicity of the texture, small thickness of the film, modal properties of the guided waves and their confinement to the film. Also we investigate the effect of the incident angle on the absorption enhancement. Our calculations provide tighter bounds for the absorption enhancement but still significant improvement is possible. Our explanation of the absorption enhancement can help better exploitation of the guided modes in thin film devices.

## 1. Introduction

Enhancing light absorption in solar cells has been a topic of research for decades. The relatively weak optical absorption of silicon at long wavelengths necessitates the application of methods to trap light inside the cell. One might thus benefit from different types of optical resonances in solar cells [1-4]. Specifically, guided mode excitation has attracted a lot of attention in recent years [3-6]. Due to their confined nature, guided modes have been regarded as a promising tool to enhance absorption in the solar cells and the extent to which they can increase light absorption in solar cells has been of interest [4-8].

Guided modes are solutions of Maxwell equations in the optical system which, ideally, do not exchange energy with the outside environment. For this, they are sometimes called "trapped modes" [8]. To excite them in a multilayered solar cell stack, the interfaces of the device must be changed from planar into textured either randomly, periodically or a combination of both [9-15].

To excite the guided modes, the following conditions need to be met [16].

1. The guided mode exists: The transverse resonance condition (TRC) of the multilayered stack should be satisfied. This means that the wave can make constructive interferences along the thickness of the guide. Approximately, this is equivalent to the condition that the resonances are solutions of the dispersion of the multilayer with flat interfaces. The approximation comes from the difference between the dispersion of the flat and the textured structure. Figure 1 shows schematically a slab waveguide textured using a one-dimensional (1D) periodic pattern. The sinusoidal curves represent symbolically change of the wave phase along the longitudinal and the transverse directions. The green curves correspond to the cases where the phase change in one round trip along the film thickness is a multiplicand of $2\pi$ i.e. where TRC is satisfied.

2. The guided mode gets excited: Periodicity must be considered by satisfying the Bragg condition. For 1D gratings and under normal illumination, this is equivalent to the condition

$$k_{\parallel} = 2m'\pi/\Lambda. \quad m': \text{integer}, \Lambda: \text{grating period} \tag{1}$$

where $k_{\parallel}$ is the propagation constant of the guided mode. The red curves in Figure 1 show the cases where phase change along the periodicity direction is a multiplicand of $2\pi$, i.e. where Bragg condition is satisfied.

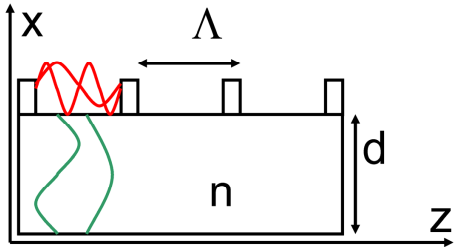

**Figure 1- (Color online) Schematics of a slab with a 1D periodic texture. The field phase variations are symbolically shown with sinusoidal patterns. The green and red curves correspond to the cases where TRC and Bragg condition are satisfied respectively.**

The two conditions above imply the existence of a discrete spectrum of the distribution of the wave-vectors and the wavelengths corresponding to guided modes. However, if $d \gg \lambda$, where $\lambda$ is the incident wavelength, the resonant wavelengths can be treated as a continuum. Similarly, if $\Lambda \gg \lambda$, the wave-vectors ($k_{\parallel}$) of the resonances can be imagined continuous. The two latter approximations have been used by different authors to find the ultimate light trapping limit [4-8]. The most well known limit was proposed by Yablonovitch *et al.* [7] and is equal to $4n^2$ where n is the refractive index of the absorber. They considered that the texture acts as a Lambertian scatterer and it scatters light into a continuum of angles (continuity of k) since they wanted to model random interfaces. Also, they assumed a thick slab of material (continuity of $\lambda$). Stuart *et al.* [8] derived the upper limit for thin absorber films but still for random textures (equivalent to large periods). Since in thin films, a guided mode is not totally confined to the structure, their limit is lower than $4n^2$. Yu *et al.* proposed the limit for thick devices with grating couplers [5, 6]. Their results tend to the $4n^2$ limit as the period of the grating is increased but over a considerable wavelength range, they surpass $4n^2$. Their calculations show that the maximal enhancement which can be obtained for thick cells is $2\pi n$ and $8\pi n^2/\sqrt{3}$ for 1D and 2D gratings respectively. Such high values are obtained for only a single wavelength. Yu *et al.* have also considered the case of a thin film that supports only a single mode. However, they assumed a continuum of resonances which is valid only for large period gratings or random textures [6]. Recently, we extended their model to thin film devices with grating couplers at the scale of the wavelength [4]; i.e. we used none of the simplifying assumptions $\Lambda \gg \lambda$ or $d \gg \lambda$.

In this work, we focus on the enhancement provided by the guided modes in periodically textured thin film solar cells and we extend the previous works on the upper limit of light trapping [4-8]. By using a coupled mode approach and investigating its underlying physics, we study the angular dependence of the limit of light absorption enhancement. Our model is distinguished from calculations of Yu *et al.* for thin films [6] by simultaneous consideration of both the thin film and the wavelength-scale grating texture. Also, our calculations can handle multiple modes in the thin film as well as a single mode in contrast to the model of Yu *et al.* for thin films. Recently, we considered the discrete nature of resonances and we showed that due to the presence of a discrete spectrum of modes, it is possible to reach very high absorption enhancement over limited wavelength ranges under normal incidence [4]. In this manuscript we extend the analysis to wide wavelength ranges and oblique illumination. Besides, we take into account the modal structure of the thin film absorber to weight the guided modes based on their impact on the absorption. For the ease of calculation, we apply our model to a dielectric slab with refractive index n=4 and thickness d= 200 nm embedded in air, but we have observed similar results for a complete solar cell stack which will be published elsewhere. The refractive index and the thickness are selected such that the film resembles a thin film amorphous silicon solar cell. While presenting the temporal coupled mode theory, we show that the enhancement factor depends on how much the energy of the guided mode is confined to the thin film absorber. This confinement is characterized with the "energy overlap" which is defined here as the fraction of the electromagnetic energy of the mode which is spread over the waveguide. We also demonstrate that if a high index cell is used e.g. a silicon solar cell, almost all energy of the incidence field occurs to be inside the film regardless of the incidence angles. Therefore, we can normally neglect the impact of the energy overlap; however, we do not use this simplification in our calculations in this paper.

This manuscript is organized as follows. First, in section 2, we describe the coupled mode theory. In section 3 we find the limit of absorption enhancement under normal illumination for a thin film solar cell using 1D grating coupling and we extend the calculations to the case of oblique incidence in section 4. The impact of the guided mode energy confinement to the cell is mainly discussed in the appendices to simplify reading the manuscript.

## 2. Overview of temporal coupled mode theory

We start with describing the impact of a resonance on the absorption enhancement. Time evolution of a single resonance with amplitude $a$, resonant angular frequency $\omega_0$, coupling $\gamma_e$ to the external radiation source $S$ and the internal loss rate of $\gamma_i$ can be described using temporal coupled mode theory [6, 17]

$$da/dt = \left[ j\omega_0 - (N\gamma_e + \gamma_i)/2 \right] a + j\sqrt{\gamma_e} S. \tag{2}$$

where $j = \sqrt{-1}$. The resonant system is assumed to have one input port and *N* output ports with the same coupling to the resonant mode; therefore, the external coupling rate $\gamma_e$ is multiplied by *N*. Since guided modes in solar cells are resonant phenomena, the latter equation can describe their temporal evolution. The coefficients $\gamma_e$ and $\gamma_i$ are assumed to be independent of the wavelength for simplicity.

It has been shown in [6] that the absorption in an absorber with the absorption coefficient $\alpha$ can be enhanced by *M* guided modes at most by a factor of

$$F = 2\pi M \gamma_i / (\alpha d \Delta \omega N) \quad (3)$$

provided that the resonances are narrowband compared to the spectral range of interest, $\Delta \omega$, and the absorber is weakly absorbing. The variable *F* in Eq. (3) is referred to as the "absorption enhancement factor" in the literature [6]. The reference of the absorption for this calculation is the single pass absorption over the same thickness as the film which is approximately equal to $\alpha d$.

In Eq. (3), the relation between $\alpha$ and $\gamma_i$ is worth investigation. If we assume that the optical wave observes the solar cell as a bulk material, the bulk approximation

$$\alpha = n \gamma_i / c \quad (4)$$

can be used [5, 6] where *c* is speed of light in air. However, Eq. (4) does not hold for thin films [4] since $\gamma_i$ is related to the "whole structure" whereas $\alpha$ describes the absorption coefficient only for the "bulk" of the absorber. Here, we find the "effective absorption coefficient" of the complete structure, say $\alpha_{wg}$, such that it can be expressed versus $\gamma_i$ analogous to Eq. 4. In general, the evolution of the amplitude $a$ of a guided mode resonance can be described in the temporal or the spatial domain

$$da/dt = (j\omega - \gamma/2)a + ...,$$
$$da/dz = (jk_\| - \alpha'/2)a + ... \quad (5)$$

where $\gamma$ and $\alpha'$ represent the photon loss rate and the damping factor, $z$ is the propagation direction along the waveguide and the source terms are not shown in Eq. (5). Both $\gamma$ and $\alpha'$ include internal loss and external coupling effects. At the absence of external coupling, $\alpha' = \alpha_{wg}$ and $\gamma = \gamma_i$ and one might conclude from Eq. (5) that

$$\frac{dz}{dt} = v_p = \frac{\omega}{k_\|} = \frac{\gamma_i}{\alpha_{wg}} = \frac{c}{n_{wg}}. \quad (6)$$

where $v_p$ is the phase velocity and $n_{wg}$ is the "effective refractive index" of the waveguide. Hence, the loss rate $\gamma_i$ is related to $\alpha_{wg}$ via

$$\alpha_{wg} = n_{wg}\gamma_i/c. \quad (7)$$

Eq. (6) is not strictly correct since it considers the phase index $n_p = k_\|/k_0$ ($k_0$ being wave vector of light in air) as the effective refractive index of the guide while the group index $n_g = \partial k_\|/\partial k_0$ provides a better approximation of the effective refractive index. This can be explained by considering a guided mode travelling inside the slab over an infinitesimal distance $dz$. The wave energy goes under a decay equal to $e^{-\alpha dz}$ in space domain or equivalently $e^{-\gamma_i dt}$ in time domain where $dt$ is the time which the wave takes to travel the distance $dz$. Since energy is guided in the waveguide at a speed equal to the group velocity, $v_g$ [16], the infinitesimal time and space elements can be related via $dt = dz / v_g$. The paradox of the latter conclusion and Eq. (6) can be explained by noting that Eq. (6) is a slowly varying approximation of the resonant mode evolution, so, it only considers the phase variations.

To obtain the link between $\alpha$ and $\gamma_i$ we still need to find the relation between $\alpha$ and $\alpha_{wg}$ since Eq. (7) links $\alpha_{wg}$ and $\gamma_i$. The ratio $\alpha_{wg} / \alpha$ is equal to the energy overlap $\eta$. This is shown for a slab in appendix A as a proof of principle. So, Eq. (3) can be rewritten as

$$F = \frac{2\pi c}{n_{wg} d \Delta \omega} \frac{M}{N} \eta. \tag{8}$$

It is explained in appendix B that for high index films, almost the whole energy of the mode will be inside the film. So, most of the times $\alpha \approx \alpha_{wg}$ ($\eta \approx 1$). To be more accurate, we do not apply this simplification to our calculations, i.e. we consider the impact of energy overlap.

In Eq. (8), the parameters $M$, $N$ and $n_{wg}$ and $\eta$ need to be calculated. The number of diffraction orders, $N$, is obtained by counting the integers $m$ which satisfy $|2\pi m / \Lambda| \leq k_0$ where $k_0$ is the wave vector of light in air. The number of guided modes, $M$, is calculated by considering the number of intersections of the device dispersion diagram (condition 1) and the lines $k_\parallel = 2\pi m' / \Lambda$ (condition 2) in the interval $\Delta \omega$ (equivalent to energy interval $\Delta E$). Figure 2 shows the dispersion diagram of the slab (n = 4 and d = 200 nm) under TE polarized illumination, i.e. the case where the electric field is normal to the incidence plane. For simplicity, this waveguide is used for our calculations throughout this paper but our method can be generalized to complete solar cells consisting of several layers. We assume that the slab is textured into a 1D grating shape with period $\Lambda = 500nm$ and we consider TE polarized illumination. The TM polarization can be treated analogously. The energy interval of interest is equivalent to wavelengths from 600 to 1200 nm since amorphous silicon cells need light trapping mainly in the range from 600 to 800 nm and the different types of crystalline silicon cells e.g. microcrystalline, monocrystalline, etc. can be enhanced by light trapping mainly in the range from 800 to 1200nm. The vertical lines correspond to different diffraction orders m=1,2,3 for normal incidence. The allowed excitations inside the wavelength range of interest are marked with circles. For each one of the allowed excitations, a specific $n_{wg}$ and $\eta$ apply.

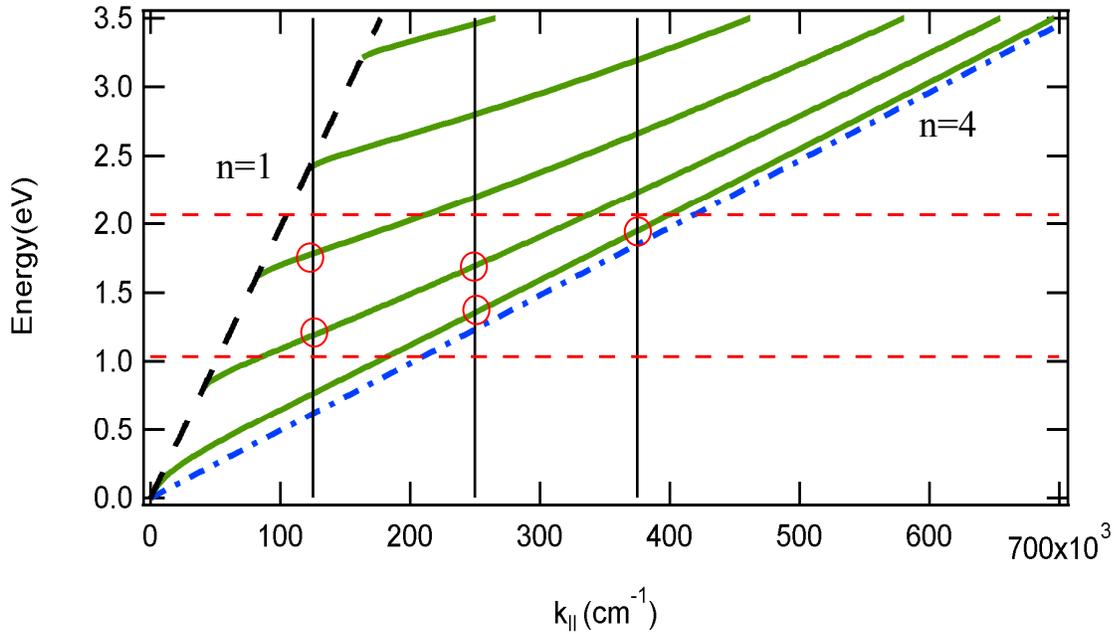

**Figure 2- (Color online) Dispersion of a film with refractive index n=4 and thickness d=200 nm in TE polarization. The guided modes (green curves) occur between the light line of air (dashed black) and dielectric (dotted dashed blue). The periodic texture excites the guided modes where they satisfy Bragg condition (vertical lines). The dashed horizontal lines correspond to 600 and 1200 nm. Resonances are illustrated by the red circles.**

Eq. (8) shows that the application of the phase index as the effective refractive index of the film can result in overestimation of the enhancement factor $F$ since $n_g > n_p$. Still, the phase index can provide an approximation of the effective index to be used in Eq. (8), especially in cases where the definition of the group index is not trivial. For example, if absorption is included in the slab, the guided modes deviate from extremely sharp spectral features into broadband resonances. Hence, there might be points in the dispersion diagram, i.e. in $(k_\parallel - \omega)$ plane, which are not on the guided modes but still occur in their close vicinity so they feel the impact of the broad guided mode resonance. For such cases and especially if more than one guided mode is influencing the absorption, a certain group index is hard to define but a unique phase index can be found to approximate $n_{wg}$. Anyway, since in this manuscript we deal with an ideal slab, we consider the group index as the effective refractive index of the film.

The presence of $n_{wg}$ instead of $n$ in the denominator is a major difference between Eq. (8) and the calculations of Ref. [6] for thick solar cells. Of course in Ref. [6] the case of a thin film solar cell has been investigated as well in which modal properties have been considered, however, there are at least two main differences between their method and our approach. First, they use a continuum model to count the resonances supported by the grating texture (Eq. (11) in [6]). This can be a good approximation for large periods but its applicability is questionable when dealing with wavelength-scale gratings since the resonances might be well distinguished in this case. Second, their model for thin films is appropriate to treat a single mode film since they count the resonances in the two dimensional $k_{yz}$ space but our model can treat multiple modes since we count the resonances in the three dimensional $k_{xyz}$ space.

In summary, by using the group index as the effective guide refractive index and applying the appropriate energy overlaps, we weight the impact of resonances based on where they appear in the dispersion diagram of Figure 2; hence, we take into account the modal properties of the waveguide.

## 3. Limit of absorption enhancement under normal incidence

Figure 3 shows the enhancement factor $F$ of the slab (n=4, d=200 nm) as a function of $\Lambda/\lambda$ for two wavelength ranges from 600 to 800 nm and from 800 to 1200 nm. $F$ is calculated for a certain grating period ($\Lambda$) in each of these two spectral ranges over tiny wavelength subintervals to consider the dispersion diagram point by point. Then, we averaged $F$ over the considered spectral range. The value $\Lambda/\lambda$ is subsequently obtained by dividing the period to the central wavelength of each interval. This gives use one point in the graphs of Figure 3. To complete the graph we need to change the period and repeat this procedure. Due to consideration of the group index as the effective refractive index of the film, the enhancement factors that we obtain are smaller than the limits suggested by Yu *et al.* for thick layers regardless of polarization and wavelength range. Similar results can be obtained for a 2D grating pattern and we have found smaller values than the 2D limit $4\pi n^2$ for square geometry 2D gratings. The only difference from the 1D case is that the counting procedure is more complicated. Note that the wavelength intervals used in our simulations are wide, however, over small wavelength ranges, higher values of *F* are obtainable.

Another important observation is that by changing the period of the grating ($\Lambda$), $F$ follows an oscillating trend [5, 6]. This type of behavior results from the balance between two different phenomena; on one hand increasing the period leads to appearance of more diffraction orders (N) and hence, reduction of *F* via Eq. (7). On the other hand, it increases the number of excited modes (M) which in turn, increases *F*. Such oscillations are observed in all our results as well as the prediction of Yu's model c.f. the black curve.

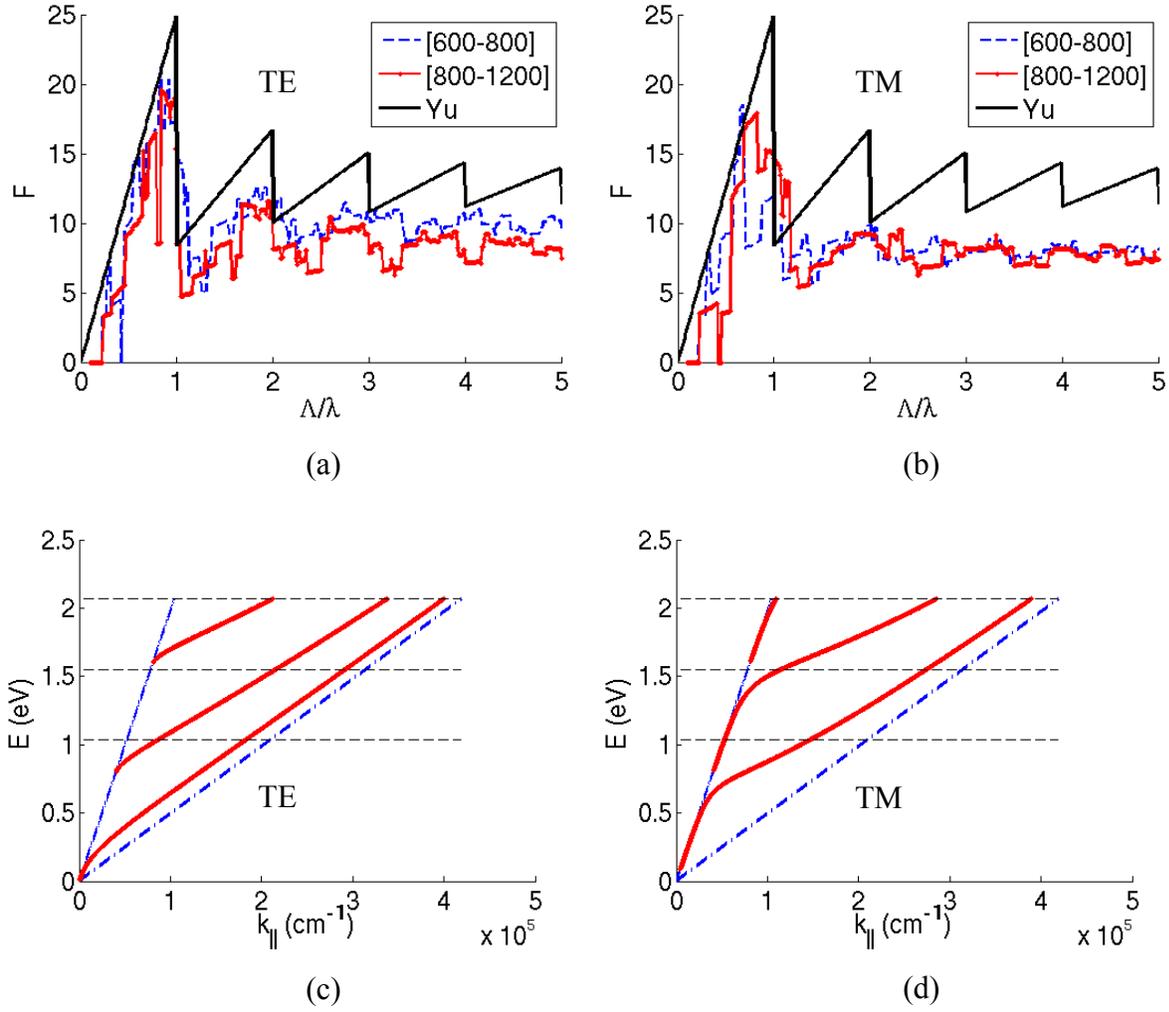

**Figure 3-** (Color online) (a) The enhancement factor introduced by 1D gratings in TE polarization versus the normalized period ($\Lambda/\lambda$) for the slab (d=200 nm, n=4). Dashed: Wavelength range from 600 to 800 nm. Solid with markers: Wavelength range from 800 to 1200 nm. Bold solid: Yu's model (thick absorber), (b) Same as (a) but for TM polarization, (c) Dispersion diagram of the slab for TE polarization, (d) same as (c) for TM polarization. The horizontal dashed lines in (c) and (d) correspond to the wavelengths 600, 800 and 1200 nm.

Figure 3 (c) shows the dispersion plot of the slab in TE polarization. The vertical dashed lines represent the boundaries of the wavelength ranges of interest, i.e. from 600 to 800 nm and from 800 to 1200 nm. In the high energy range ([600-800] nm) there are three guided modes but at low energy range ([800-1200] nm), there are only two modes. This is the reason why in Figure 3 (a) TE polarized light can result in higher $F$ in the high energy range compared to the low energy range. In TM polarization - Figure 3 (d)-, the highest order mode is very close to the light line of air and it is shown in Appendix B that a major part of its energy extends over the region outside the slab. Therefore, there remain two TM modes which occur in both energy ranges and there is not a significant difference between the *F* of the two energy ranges in TM polarization as depicted in Figure 3 (b).

It is worth mentioning that due to Eq. (8) huge enhancement factor are expected close to the light line of air regardless of taking the group index or the phase index since both of these quantities takes their lowest values there. This happens for example for the second TM mode in Figure 3 (d). However, this advantage might be compromised by the low confinement of the mode to the guide in some cases as the third TM mode in the last example. Nevertheless, one should be able to find a region of optimized conditions close to the light line of air.

## 4. Absorption limit under oblique illumination

In this section we calculate the absorption enhancement limit provided by the guided mode excitation at arbitrary angles. For simplicity, we consider in-plane incidence over the slab (n=4, d=200nm) as depicted schematically in Figure 4 (a) and we investigate the impact of a 1D texture. By in-plane incidence we mean that the incident beam and the reflection orders are all in x-z plane. The incident angle $\theta$ affects the calculations only by changing the Bragg condition

$$k_{\parallel,m}' = k_{\parallel,m} + k_0 \sin\theta \tag{9}$$

where $k_{\parallel,m}$ and $k_{\parallel,m}'$ show the parallel component of the wave vector of the *m*-th diffraction order at normal and oblique incidence respectively and $k_0$ is the wave vector of the light in the air. Figure 4 (b) shows the variation of the energy (equivalently the resonant frequency) of positive and negative orders introduced by changing the incident angle. The vertical and tilted dashed lines correspond to the Bragg conditions under normal and oblique illumination both for $\Lambda = 500\ nm$. For simplicity, only one mode of the film and two orders ($\pm 1$) are demonstrated between the light lines of the air and the dielectric $n=4$. By a positive change of $\theta$ from zero, the positive orders go farther from the origin ($k_\parallel = 0$) but the negative orders get closer to it. This is shown in Figure 4 (b) where the dashed lines correspond to the case in which $\theta = 20°$ has been changed. As Eq. 9 suggests, both positive and negative orders go under a shift equal to $k_0 \sin\theta$ which means that they do not excite the same guided mode of the film at the same energy any more. This asymmetry in satisfying the resonance condition is a natural result of the breaking symmetry provoked by changing the incident angle.

Figure 4 (c) and (d) show the angular behavior of $F$ for the slab (n=4, d=200nm) under TE and TM polarized light over the wavelength range from 600 to 800 nm and Figure 4 (e) and (f) are the same as Figure 4 (c) and (d) for the interval from 800 to 1200 nm. As explained in appendix B, the impact of energy overlap is negligible except very close to the line of the air; still we consider it in our calculations. At normal incidence ($\theta = 0°$), $F$ follows the same sawtooth trend as observed in Figure 3. However, as the incident angle is increased, the peaks of $F$ are moved in an ordered way in both polarization directions.

The peaks of $F$ move in the period-angle plane on two series of curves with positive and negative slopes. These two groups of curves are associated with the positive and negative diffraction orders. Increasing the incidence angle results in higher resonant energy for positive orders as shown in Figure 4 (b). This means that for a fixed grating period, the normalized period ($\Lambda/\lambda$) will become larger too. Therefore, the curves with positive slopes in Figure 4 correspond to the appearance of the positive diffraction orders. Similarly, it is concluded that the curves with negative slope represent the negative diffraction orders. Figure 4 shows that relatively high enhancement factors can be obtain at angles other than normal. This is a direct result of the asymmetric shift of the resonant energies of the positive and negative orders under oblique illumination.

Figure 5 shows the enhancement factor of the 1D grating discussed in Figure 4 for only two incident angles of 20° and 60° and for the first wavelength interval i.e. from 600 to 800nm. It is observed that the smaller incident angle permits higher $F_{max}$ at periods close to the incident wavelength. However, for larger periods, there is no significant difference between the two angles. This is a general conclusion and agrees with the intuition that at the limit of large period, the grating should resemble a Lambertian scatterer [6].

Altogether, the incident angle affects the absorption by changing the number of diffraction orders. Close to the normal incidence, absorption can be more dramatically enhanced. Since solar illumination occurs at a limited range of incidence angles, it is possible to keep the condition of normal incidence at least to some extent. Based on the wavelength range which affects the performance of the cell, periods can be found which can enhance light more drastically.

It is worth mentioning that in all the calculations in this manuscript and also in the prior research [4-8], the impact of the coupling strength is still missing. It is expected that by considering the coupling efficiency, $F$ drops to values smaller than the ones presented here and maybe all of the other mentioned works [4-8]. Also, other loss mechanisms such as parasitic absorption, imperfect collection of photo-generated carriers and reflection from the top interface of the cell would lead to a decrease in the achievable photocurrent.

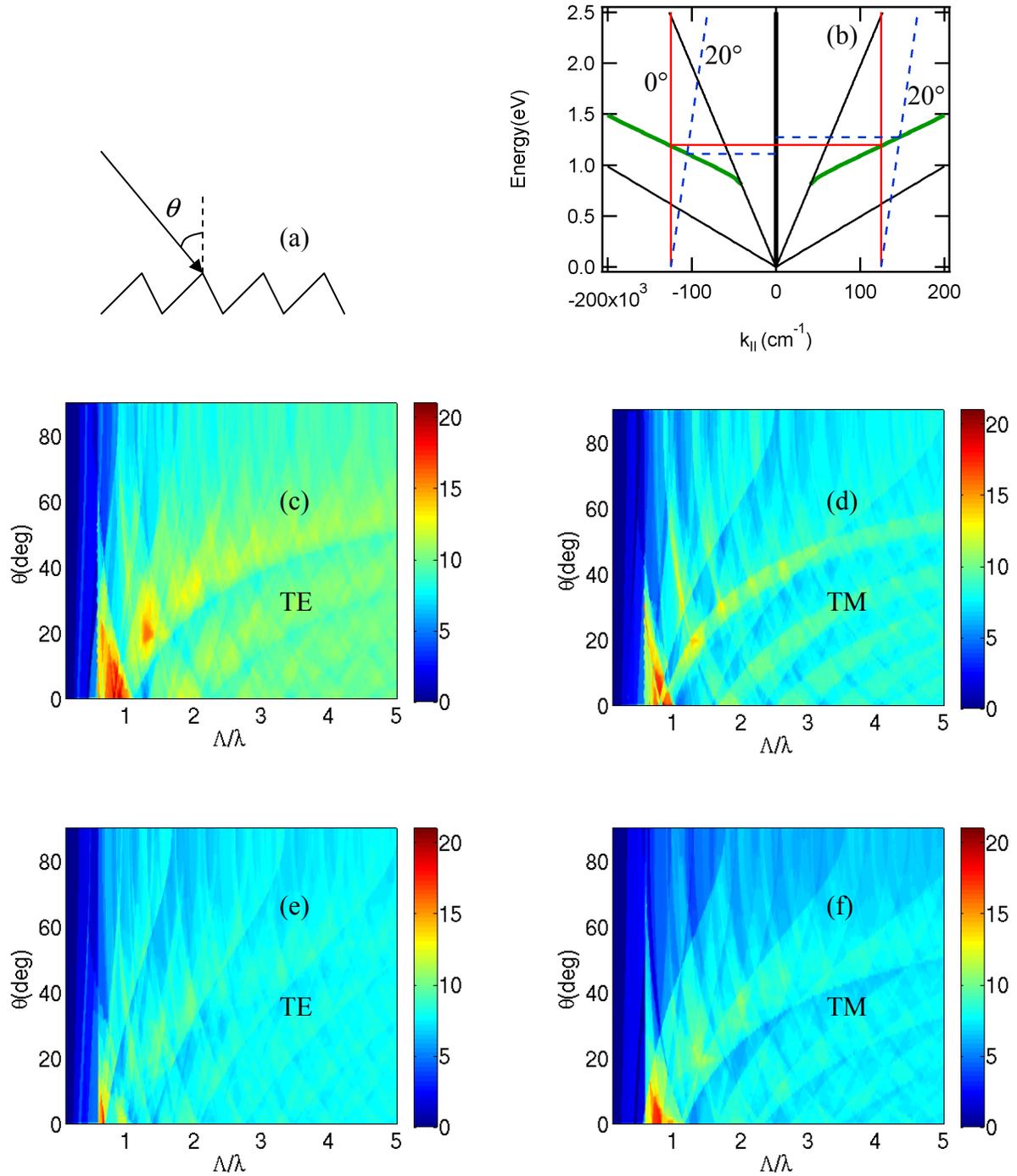

**Figure 4-** (Color online) (a) Schematic view of the grating under in-plane oblique incidence. (b) shift of the resonant energy of the positive and negative orders due to the change of the incident angle from 0° to 20°. (c) Angular dependence of the *F* under TE polarized illumination over the wavelength range [600-800] nm. (d) same as "c" for TM polarized illumination. (e) same as "c" for the wavelength range [800-1200] nm. (f) same as "e" for TM polarized illumination.

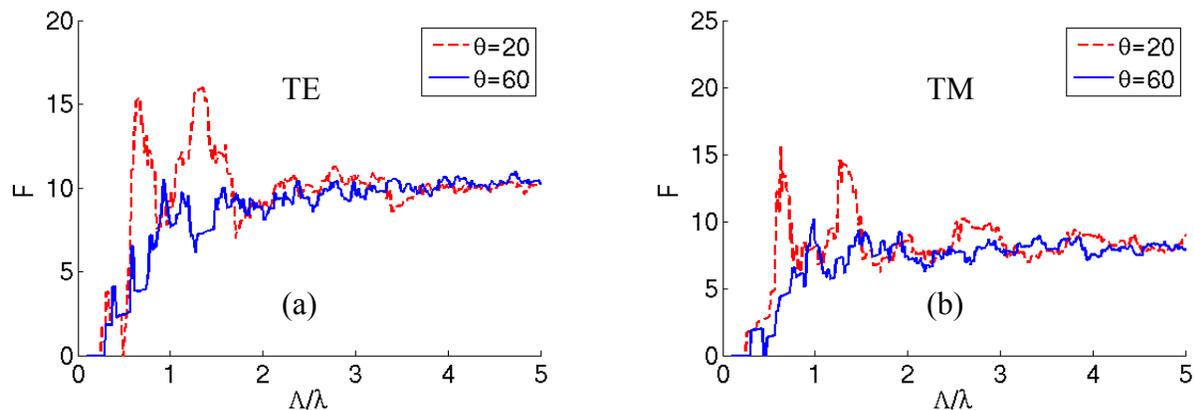

**Figure 5-** (a) *F* under TE polarized illumination over the wavelength range [600-800] nm for the incident angles of 20° and 60°. (b) same as "a" but for TM polarization.

## 5. Conclusions

In summary, we study light absorption enhancement using guided modes at normal and oblique incidence. We include simultaneously the small thickness of the cell and wavelength-scale periodicity of the texture in the coupled mode approach previously introduced by Yu *et al.* [6] to find the upper bound of the absorption enhancement in thin film silicon solar cells. In Ref. [6], the authors did not consider these two at the same time so they were able to approximate the number of resonance in the solar cell with a continuum. However, our model takes into account the discrete nature of the diffraction phenomena and the guided modes in thin film solar cells. In our calculations we include also the effect of confinement of the guided mode energy to the thin film absorber as previously suggested [6, 8]. Our results are obtained over two wide wavelength ranges: from 600 to 800 nm and from 800 to 1200 nm to meet the light trapping conditions of amorphous and crystalline silicon solar cells. We show that by benefiting from the discrete spectrum of the guided modes and their modal properties in thin films, it is possible to enhance absorption significantly, however, the absorption enhancement limit is lower than the values predicted by Yu. *et al.* [6]. We also investigate the effect of incident angle. Our calculations show that increasing the incident angle does not necessarily reduce the absorption enhancement factor. We attribute this to the asymmetric shift of the resonant frequencies corresponding to positive and negative Bragg orders.

## Appendix A: The absorption coefficients

The wave loses energy with the effective absorption coefficient $\alpha_{wg}$ as it is guided along the film. If the whole space is filled with the absorber, this absorption coefficient will be equal to $\alpha$. In a film with finite thickness, energy will be distributed over both the device and the outer space with the energy overlap $\eta$. The wave energy after passing a length $dz$ is

$$\mathrm{E}(dz) = \mathrm{E}(0) e^{-\alpha_{wg} dz} = \mathrm{E}(0)\left[\eta e^{-\alpha dz} + (1-\eta)\right] \tag{A1}$$

Taylor expansion of (A1) leads to $\alpha_{wg}/\alpha = \eta$.

## Appendix B: The energy overlap

At normal incidence, the electric field profile of even and odd modes can be easily obtained [16].

$$E_e = \begin{cases} \cos(k_x x)/\cos(k_x d/2) & |x| \le d/2, \\ \exp(-\alpha(|x| - d/2)) & |x| > d/2. \end{cases} \tag{B1-a}$$

$$E_o = \begin{cases} \sin(k_x x)/\sin(k_x d/2) & |x| \le d/2, \\ \exp(-\alpha(x - d/2)) & x > d/2, \\ -\exp(-\alpha(-x + d/2)) & x < -d/2 \end{cases} \tag{B1-b}$$

where $x$ refers to the direction normal to the interfaces and $E_o$ and $E_e$ refer to electric field of the odd and the even modes respectively. The energy overlap can be obtained via the electric field profiles.

$$\eta = \left(n^2 \int |E_{slab}|^2 \, dx\right) \Big/ \left(n^2 \int |E_{slab}|^2 \, dx + \int |E_{air}|^2 \, dx\right) \tag{B2}$$

At non-perpendicular incidence, TE and TM polarizations must be distinguished. As in the case of normal incidence, the guided modes can be classified into odd and even in each polarization.

In TE polarization, the electric field is parallel to the slab boundary and the electric field of the modes can be expressed with Eq. (B1-a,b). Therefore, the energy overlap $\eta$ can be obtained similarly to the case of normal incidence. Figure 6 (a) and (b) show the energy overlap versus phase index $n_p$ and photon energy E for the slab ($d = 200 nm$, $n = 4$) under oblique incidence. Since $\eta$ is the same as the case of normal incidence, Figure 6 (a) and (b) can also be used for normal illumination of light. The energy overlap is very close to unity almost everywhere except near the light line of air ($n_p \approx 1$). The guided modes corresponding to each case are shown in Figure 6 with red dotted curves. In a narrow spectral range, there is an even TE mode partly in the low $\eta$ region close to the light line of air. As this mode approaches the light line of air, its group index decreases gradually, therefore, Eq. (7) predicts that the mode should provide high enhancement factors. However, this advantage is compromised by the small $\eta$ which means extension of the wave tail outside the guide and subsequently, reduction of F.

In TM polarization, the magnetic field has a form similar to the electric field in TE polarization but the electric field has two components which can be derived from the magnetic field by using Maxwell equations.

$$E_x = \frac{-j}{\omega \varepsilon} \frac{\partial}{\partial z} H_y, \tag{B3-a}$$

$$E_z = \frac{j}{\omega\varepsilon}\frac{\partial}{\partial x}H_y. \quad \text{(B3-b)}$$

By applying Eq. (B3) and using the form of the magnetic field which is already described in Eq. (B1), the electric field profiles in TM polarization can be obtained. To calculate the intensity profile, both components of the electric field must be considered.

$$I = |E|^2 = |E_x|^2 + |E_z|^2 \quad \text{(B4)}$$

After some algebraic manipulation the following intensity profiles can be obtained for the TM polarized modes.

$$I_e = \begin{cases} \dfrac{\mu_0}{\varepsilon_0 n^2}\dfrac{k_x^2 \sin^2(k_x x) + \beta^2 \cos^2(k_x x)}{k_0^2 \cos^2(k_x d/2)} & |x| \leq d/2 \\ \dfrac{\mu_0}{\varepsilon_0}\exp(-2\alpha(|x|-d/2)) & |x| > d/2 \end{cases} \quad \text{(B5-a)}$$

$$I_o = \begin{cases} \dfrac{\mu_0}{\varepsilon_0 n^2}\dfrac{k_x^2 \cos^2(k_x x) + \beta^2 \sin^2(k_x x)}{k_0^2 \sin^2(kd/2)} & |x| \leq d/2 \\ \dfrac{\mu_0}{\varepsilon_0}\exp(-2\alpha(|x|-d/2)) & |x| > d/2 \end{cases} \quad \text{(B5-a)}$$

where $I_o$ and $I_e$ refer to intensity of the odd modes and even modes respectively. The energy overlap can be obtained via the intensity profiles by applying Eq. (B2).

Figure 6 (c) and (d) show the energy overlap for the slab ($d = 200nm$, $n = 4$) under TM polarized illumination. As in the previous cases, the energy overlap is almost unity everywhere except near the light line of the air ($n_p = 1$). Figure 6 (c) shows an even TM mode which occurs only very close to the light line of air. Since $\eta$ is very small in this region, this mode cannot result in very huge absorption enhancement although its group index is small.

Altogether, the incident angle does not have a dramatic effect on the energy overlap for high index cells. Hence, one can almost always approximate $\eta \approx 1$ by considering a margin around the light line of air and due to Appendix A, $\alpha_{wg} \approx \alpha$.

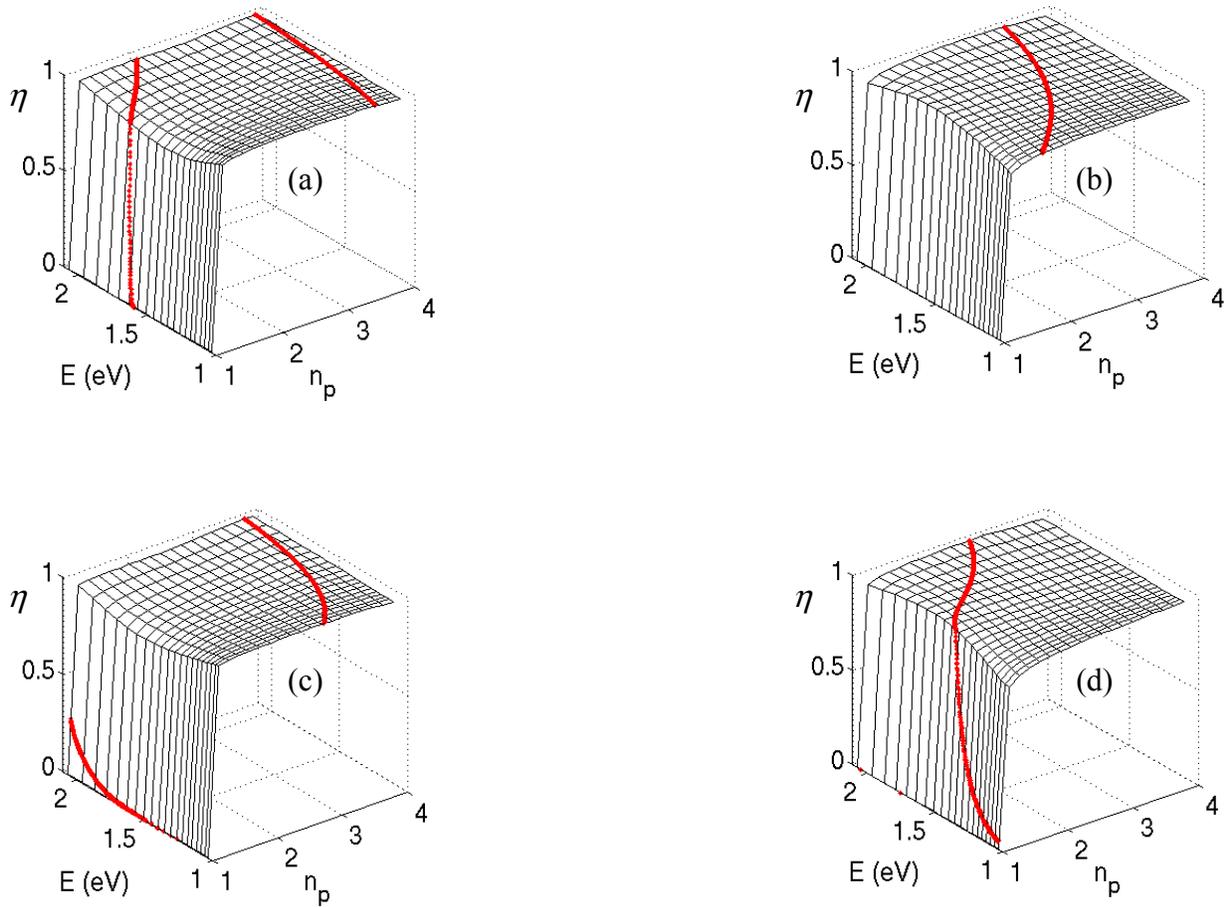

**Figure 6- (Color online) The energy overlap as a function of photon energy (eV) and phase index for (a) even modes under TE polarized light, (b) odd modes under TE polarized light (c) even modes under TM polarized light (d) odd modes under TM polarized light. The energy range corresponds to the wavelengths between 600 and 1200nm. The red dotted curves represent the guided modes corresponding to each case.**

Note that the diagrams of Figure 6 are plotted for a continuum of energy and phase index values but dispersion of the film produces a discrete spectrum. So, on the surfaces in Figure 6, only the $n_p$ and E values corresponding to the guided modes i.e. the red dotted curves should be considered.

## Acknowledgement

We acknowledge Gaël D. Osowiecki for proofreading of the manuscript. This work was supported by the Swiss National Science Foundation for funding under Project No. 200021_125177/1.

## References

1. H. A. Atwater and A. Polman, "Plasmonics for improved photovoltaic devices," Nat Mater **9**, 205-213 (2010).
2. J. Grandidier, D. M. Callahan, J. N. Munday, and H. A. Atwater, "Thin-Film Solar Cells:Light Absorption Enhancement in Thin-Film Solar Cells Using Whispering Gallery Modes in Dielectric Nanospheres," Advanced Materials **23**, 1272–1276 (2011).
3. A. Naqavi, K. Söderström, F.-J. Haug, V. Paeder, T. Scharf, H. P. Herzig, and C. Ballif, "Understanding of photocurrent enhancement in real thin film solar cells: towards optimal one-dimensional gratings," Optics Express **19**, 128-140 (2011).
4. F.-J. Haug, K. Söderström, A. Naqavi, and C. Ballif, "Resonances and absorption enhancement in thin film silicon solar cells with periodic interface texture," Journal of Applied Physics **109**, 084516 (2011).
5. Z. Yu, A. Raman, and S. Fan, "Fundamental limit of light trapping in grating structures," Optics Express **18**, A366-A380 (2010).
6. Z. Yu, A. Raman, and S. Fan, "Fundamental limit of nanophotonic light trapping in solar cells," Proceedings of the National Academy of Sciences **107**, 17491-17496 (2010).
7. E. Yablonovitch and G. D. Cody, "Intensity enhancement in textured optical sheets for solar cells," IEEE Transactions on Electron Devices **29**, 300-305 (1982).
8. H. R. Stuart and D. G. Hall, "Thermodynamic limit to light trapping in thin planar structures," JOSA A **14**, 3001-3008 (1997).
9. O. Isabella, F. Moll, J. Krč, and M. Zeman, "Modulated surface textures using zinc oxide films for solar cells applications," physica status solidi (a) **207**, 642-646 (2010).
10. C. Battaglia, C. M. Hsu, K. Söderström, J. Escarré, F. J. Haug, M. Charrière, M. Boccard, M. Despeisse, D. Alexander, and M. Cantoni, "Light Trapping in Solar Cells: Can Periodic Beat Random?," ACS nano **6**, 2790–2797 (2012).

11. J. Krč, M. Zeman, O. Kluth, F. Smole, and M. Topič, "Effect of surface roughness of ZnO: Al films on light scattering in hydrogenated amorphous silicon solar cells," Thin solid films **426**, 296-304 (2003).
12. C. Rockstuhl, S. Fahr, K. Bittkau, T. Beckers, R. Carius, F. Haug, T. Söderström, C. Ballif, and F. Lederer, "Comparison and optimization of randomly textured surfaces in thin-film solar cells," Optics Express **18**, A335-A341 (2010).
13. M. Python, E. Vallat-Sauvain, J. Bailat, D. Dominé, L. Fesquet, A. Shah, and C. Ballif, "Relation between substrate surface morphology and microcrystalline silicon solar cell performance," Journal of Non-Crystalline Solids **354**, 2258-2262 (2008).
14. P. Bermel, C. Luo, L. Zeng, L. C. Kimerling, and J. D. Joannopoulos, "Improving thin-film crystalline silicon solar cell efficiencies with photonic crystals," Optics Express **15**, 16986-17000 (2007).
15. C. Heine and R. Morf, "Submicrometer gratings for solar energy applications," Applied Optics **34**, 2476-2482 (1995).
16. T. Tamir, "Integrated optics," Topics in Applied Physics, Berlin: Springer, 1979, 2nd rev. ed., edited by Tamir, T. **1**(1979).
17. H. A. Haus, *Waves and fields in optoelectronics* (Prentice-Hall  Englewood Cliffs, NJ, 1984).

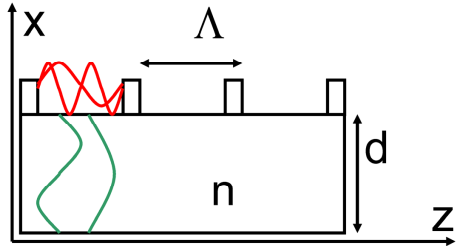

**Figure 7- (Color online) Schematics of a slab with a 1D periodic texture. The field phase variations are symbolically shown with sinusoidal patterns. The green and red curves correspond to the cases where TRC and Bragg condition are satisfied respectively.**

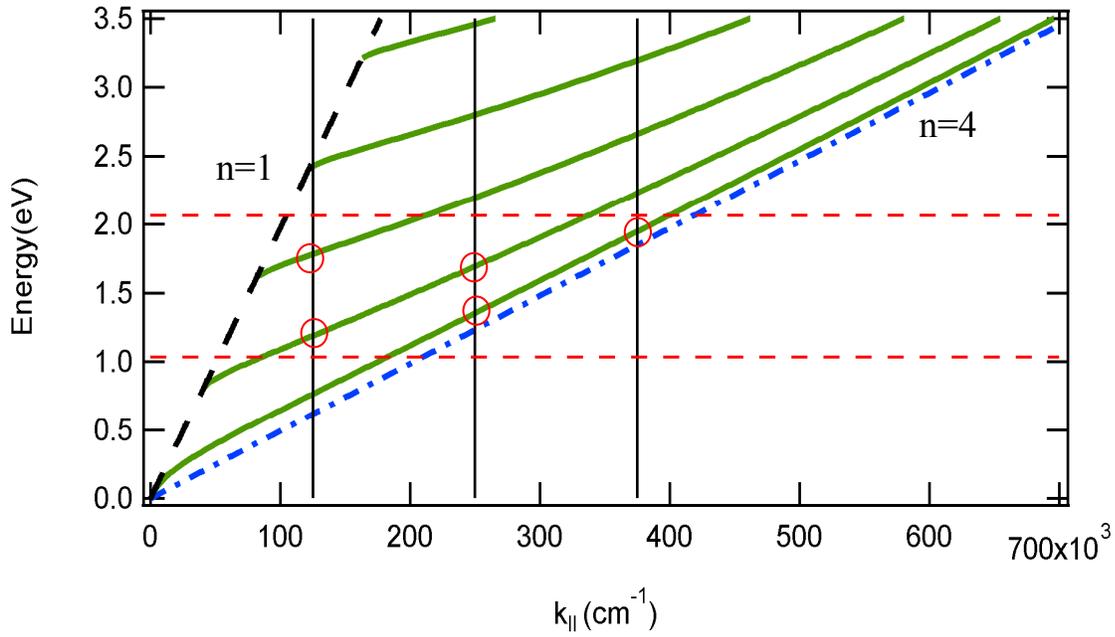

**Figure 8-** (Color online) Dispersion of a film with refractive index n=4 and thickness d=200 nm in TE polarization. The guided modes (green curves) occur between the light line of air (dashed black) and dielectric (dotted dashed blue). The periodic texture excites the guided modes where they satisfy Bragg condition (vertical lines). The dashed horizontal lines correspond to 600 and 1200 nm. Resonances are illustrated by the red circles.

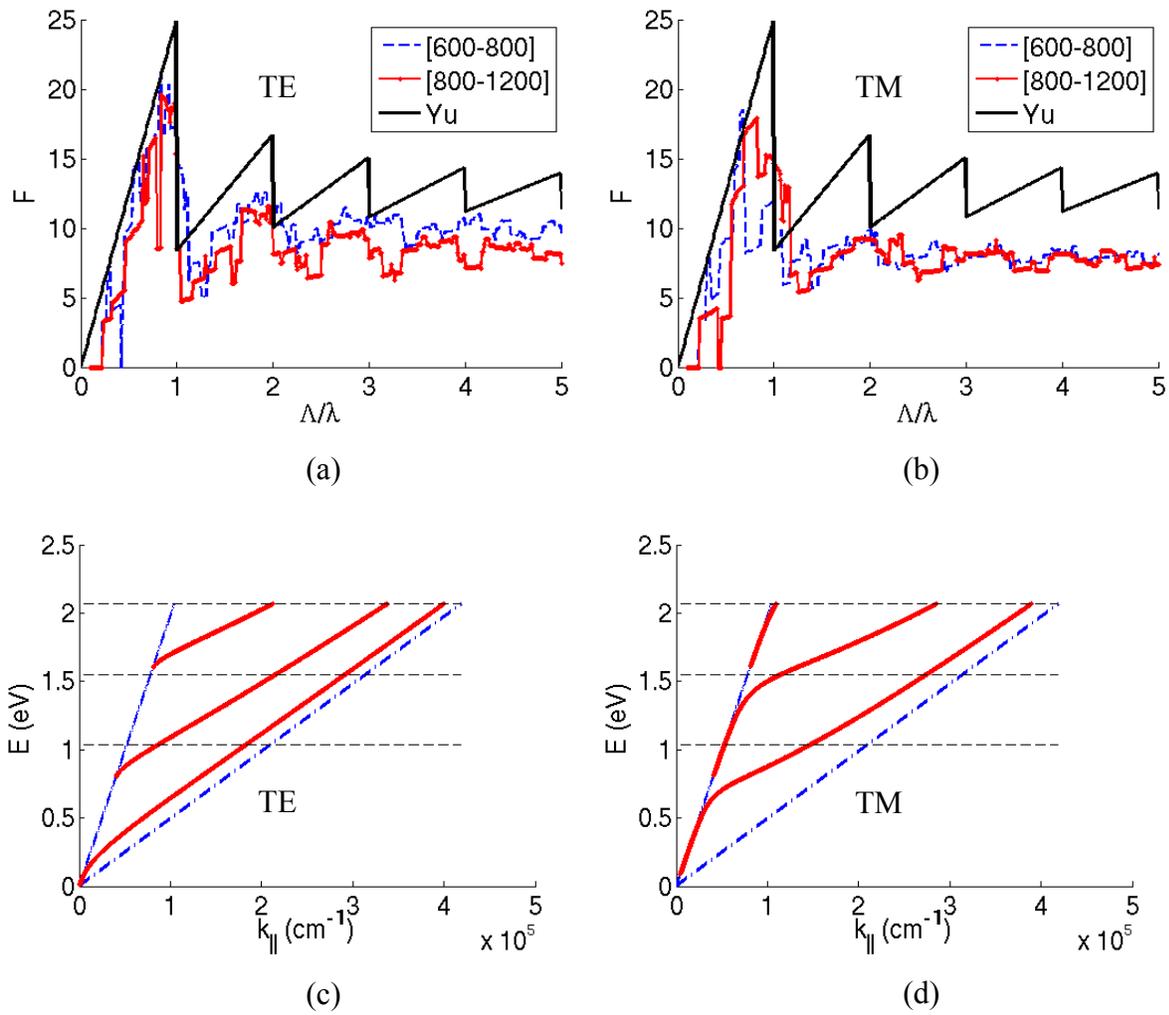

**Figure 9-** (Color online) (a) The enhancement factor introduced by 1D gratings in TE polarization versus the normalized period ($\Lambda/\lambda$) for the slab (d=200 nm, n=4). Dashed: Wavelength range from 600 to 800 nm. Solid with markers: Wavelength range from 800 to 1200 nm. Bold solid: Yu's model (thick absorber), (b) Same as (a) but for TM polarization, (c) Dispersion diagram of the slab for TE polarization, (d) same as (c) for TM polarization. The horizontal dashed lines in (c) and (d) correspond to the wavelengths 600, 800 and 1200 nm.

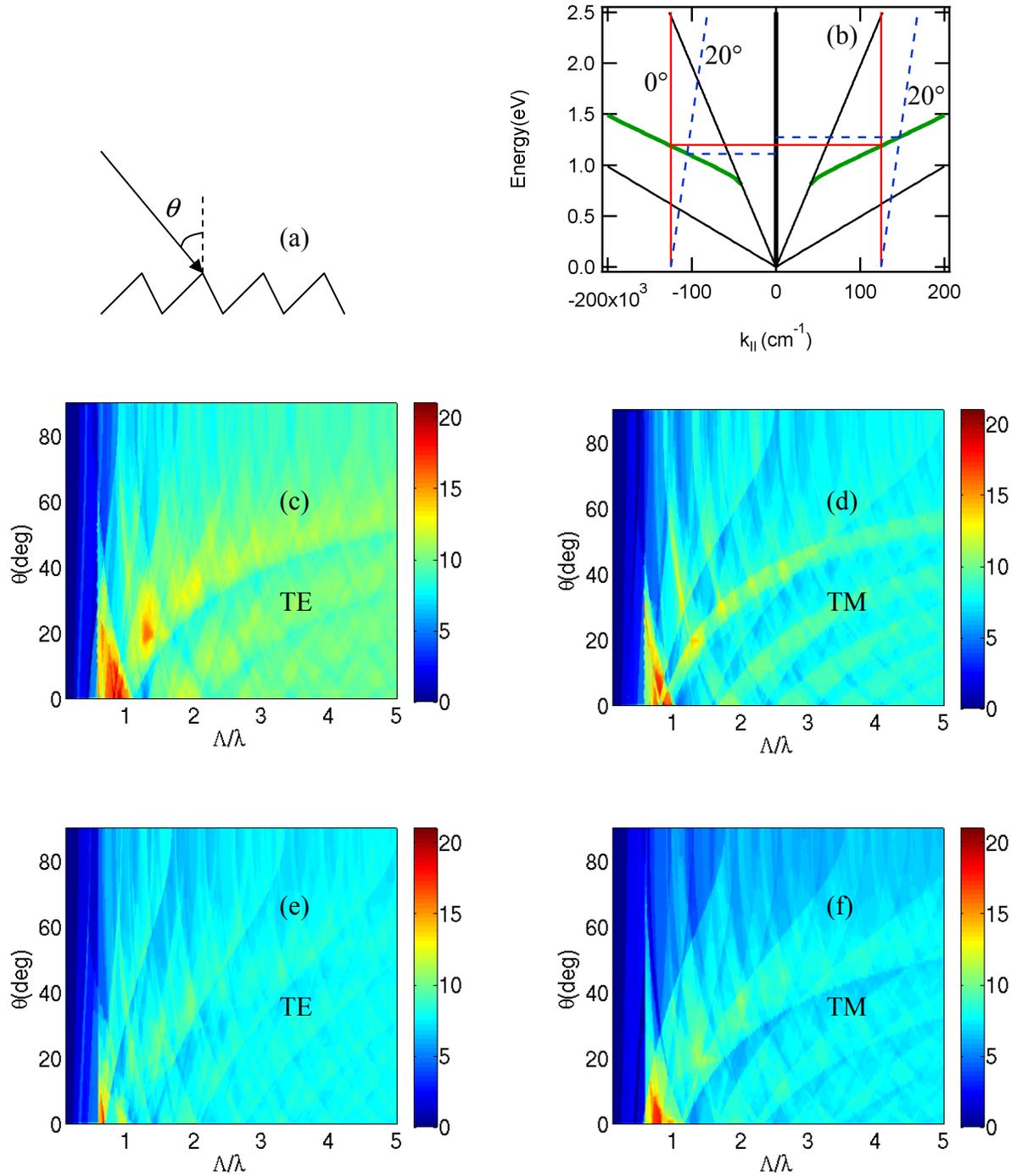

**Figure 10-** (Color online) (a) Schematic view of the grating under in-plane oblique incidence. (b) shift of the resonant energy of the positive and negative orders due to the change of the incident angle from 0° to 20°. (c) Angular dependence of the *F* under TE polarized illumination over the wavelength range [600-800] nm. (d) same as "c" for TM polarized illumination. (e) same as "c" for the wavelength range [800-1200] nm. (f) same as "e" for TM polarized illumination.

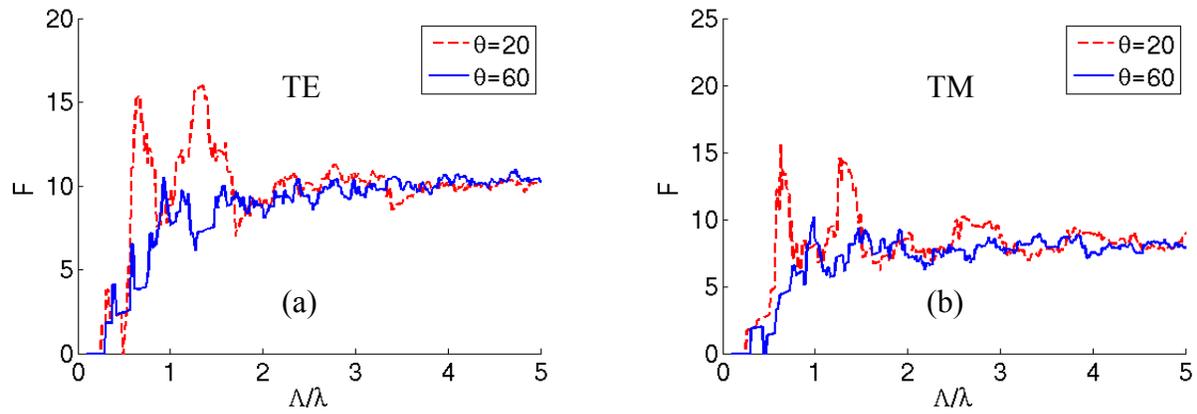

**Figure 11- (a)** *F* **under TE polarized illumination over the wavelength range [600-800] nm for the incident angles of 20° and 60°. (b) same as "a" but for TM polarization.**

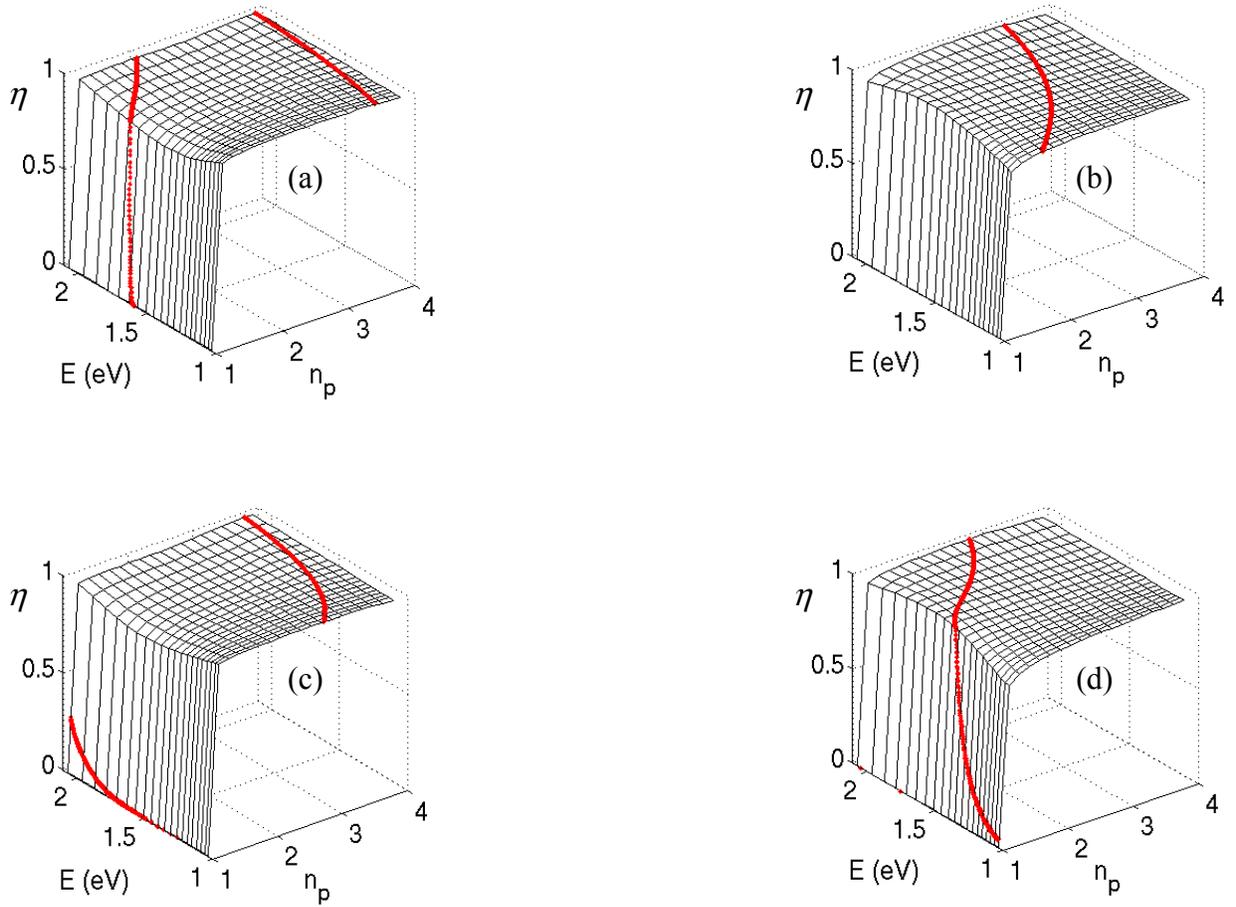

**Figure 12- (Color online) The energy overlap as a function of photon energy (eV) and phase index for (a) even modes under TE polarized light, (b) odd modes under TE polarized light (c) even modes under TM polarized light (d) odd modes under TM polarized light. The energy range corresponds to the wavelengths between 600 and 1200nm. The red dotted curves represent the guided modes corresponding to each case.**